\documentclass[letterpaper,11pt]{article}
\pdfoutput=1
\usepackage{jheppub_mod}
\usepackage{graphicx, amsmath, amssymb, amstext,xcolor,txfonts}
\usepackage{relsize, float, multirow}
\usepackage{array, epstopdf, hyperref}
\usepackage{caption, subcaption, mathtools}
\usepackage{newtxmath,ulem}

\definecolor{seagreen}{rgb}{0.18, 0.55, 0.34}

\title{Redshift dependence of FRB host dispersion measures across cosmic epochs}

\author[a,1]{Sandeep Kumar Acharya,\note{Corresponding author.}}
\author[a,b,c]{Paz Beniamini}

\affiliation[a]{Astrophysics Research Center of the Open University, The Open University of Israel, Ra'anana, Israel}
\affiliation[b]{Department of Natural Sciences, The Open University of Israel, P.O Box 808, Ra’anana 4353701, Israel}
\affiliation[c]{Department of Physics, The George Washington University, 725 21st Street NW, Washington, DC 20052, USA}

\emailAdd{sandeepa@openu.ac.il}
\date{\today}

\abstract{ We constrain the redshift dependence of (rest frame) host galaxy dispersion measures of localized Fast Radio Bursts (FRBs) by assuming it to vary as a simple power law ($\propto (1+z)^{\alpha}$). We simultaneously fit $\alpha$ as well as the host dispersion measure to the data of FRBs with known redshifts. We find that \color{black} $\alpha$ between 0 to 1 is preferred depending upon our modelling choices. Current data can constrain $|\alpha|\lesssim 2$ at a 68 percent confidence interval \color{black}. Such constraints have implications for our understanding of galaxy formation and can be used to inform galaxy and large scale simulations.     }
\notoc
\begin{document}
\maketitle
\newpage

\section{Introduction}
FRBs (Fast Radio Bursts) are transients with order millisecond durations and typically observed at frequencies in the range of 100 MHz-8 GHz. The first FRB was discovered by \cite{Lorimer2007}. Thousands of FRBs have been discovered to date \citep{Petroff2016, Chime2019,Heintz2020,AKSAP2022,Law2024} while a few dozens have been identified to a host galaxy and hence have an identified redshift \citep{Chatterjee2017,Zhang2023}. Additionally, several tens of FRBs have been shown to be repeating in nature \citep{Spitler2016}. Even with this progress, we still have only a limited understanding of the origin of FRBs. For a review on FRBs, their properties, implications and possible explanations, the readers are referred to the following reviews \citep{Petroff2022,Zhang2023}.

The radio pulses from FRBs get dispersed as they travel through ionized medium along the line of sight. The time of arrival of the radio waves is proportional to DM$\cdot \nu^{-2}$ where DM is the dispersion measure which is locally given by $\int n_{\rm e}{\rm d}l$, where $n_{\rm e}$ is the electron density and ${\rm d}l$ is a line element along the line of sight, which are measured in the rest frame of the medium. The DM measured by the observer is Lorentz transformed from the local source frame by recalling that the DM transforms as the frequency. The observed DM far exceeds the Galactic contribution measured from pulsars. This means that a large fraction of the DM is contributed by propagation of the FRB signals before entering our Galaxy and makes FRBs useful probes for cosmological applications. Also, DM is straight-forward to measure and is available with high accuracy for each detected FRB. Several cosmological applications such as determining the baryon contents in the Universe, measuring the Hubble constant and probing the era of reionization have been discussed in the literature \citep{McQuinn2014,Eichler2017,ZE2018,LGDWZ2018,KL2019,Macquart2020,Beniamini2021,HR2022,James2022,WZW2022,Zhang2023,Khrykin2024,Gao2024}.  

The contribution to the observed DM is a combination of DM from the host galaxy, intergalactic medium (IGM), our own Galaxy and the surrounding halo. The IGM term captures the expansion and the baryon content of the Universe. Therefore, it is of primary interest for cosmological applications. However, generally, we do not know the individual components of the total observed DM. Therefore, one needs to have an understanding of the other components in order to extract the maximum cosmological information from given data. Among all the components, the host galaxy is, probably, the most important and complex to understand. It depends upon star formation inside the galaxy as well as baryonic feedback processes. Previous works have jointly fitted cosmological parameters such as the Hubble parameter along with the host galaxy contribution (assuming it to be redshift-independent) from the data itself \citep{Macquart2020}.
The average value of host galaxy contribution from these analyses turns out to be of the order of 50-100 pc cm$^{-3}$. 
Other works \citep{Zhang2020,Theis2024,Mo2023} have used cosmological simulations to compute the host galaxy contribution. These results show that the host galaxy contribution is typically sub-dominant compared to the intergalactic medium. \cite{Zhang2020}  reported that the host galaxy contribution increases with redshift (approximately $\propto (1+z)$ on average) though it depends sensitively on the type of galaxy. Recently, another work \citep{Kovacs2024} studied the DM evolution with redshift using the IllustrisTNG50 simulation \citep{Pillepich2019} and obtained similar results (see Fig. 4 of the reference for comparison) as in \cite{Zhang2020}. The authors attributed this change to increase in electron density inside the galaxies due to star formation. However, these cosmological simulations lack resolution at smaller scales which can potentially bias their results. A recent work \citep{Orr2024} shows that the host galaxy contribution can be dominant or be as important as the intergalactic medium. This work utilised higher resolution FIRE-2 simulations \citep{Hopkins2018} to investigate host DM contribution.        

In this work, considering the theoretical uncertainty based on cosmological simulations, we outline a method to empirically constrain the redshift dependence of the mean host galaxy contribution and its redshift evolution, characterized here as a power-law, $\propto (1+z)^{\alpha}$. The redshift-dependence of host galaxies, if any, can have important consequences for galaxy formation models and will capture cosmological relevant information such as their evolution across cosmic time. Our analysis is based on a few assumptions. We have fixed the halo contribution to a value which has been used in the literature \citep{Macquart2020,HR2022}. \color{black}We will show that our results qualitatively do not change for a different choice of halo ${\rm DM}$ \color{black}. Subject to these caveats, we find that the current data can exclude 
$\alpha\gtrsim 2$. \color{black}However, these results are sensitive to the choice of ${\rm DM}$ contribution from the IGM. \color{black}  In the future, we may rule out (or confirm) $\alpha\gtrsim 1$ which will critically test the results of \cite{Zhang2020,Kovacs2024}.  We also show that the inferred, best fit value of of cosmological parameters such as the baryon density of the Universe and the Hubble parameter are stable to a change in the index $\alpha$. We organize the paper as follows. In Sec. \ref{Sec:DM_FRB}, we discuss the various contribution to the observed DM along with their modelling and introduce our FRB sample. We discuss our procedure for computing likelihood and fitting our model to the data in Sec. \ref{sec:Likelihood}. In Sec. \ref{Sec:results}, we discuss our results and end with discussions in Sec. \ref{Sec:conclusions}.  

\section{Dispersion measure of FRBs and choice of sample}
\label{Sec:DM_FRB}
Radio waves get dispersed due to intervening ionized medium as they travel from the source to us. The observed or total DM can be expressed as,  
\begin{equation}
  \rm{  DM_{obs}=DM_{halo}+DM_{ISM}+DM_{IGM}+\frac{DM_{\rm host}+DM_{source}}{1+z}}
  \label{eq:total_DM}
\end{equation}
where $\rm {DM_{halo}}$ and $\rm {DM_{ISM}}$ are the contributions from our Galaxy and the surrounding halo. The IGM contribution comes from the large scale structure of the Universe and depends upon the assumed cosmological model. The host and source terms are contributed by the host galaxy and the immediate surroundings of FRBs respectively, which is redshifted in the observer frame. We have dropped the source term (which for most sources is expected to be subdominant) in this work. \color{black}The source contribution is only expected to be large when the source is relatively young ($t_{\rm age}\lesssim 100$yr) \citep{Piro2016,MBM2017,YZ2017}. At such a source age FRBs are likely to also have large rotation measure \citep{Hilmarsson2021,AnnaThomas2023} and persistent emission \citep{Chatterjee2017,Niu2022} (which should be correlated with large DM from this immediate environment). By taking out such FRBs, i.e. 20190520B and 20121102A from our sample, we can minimize the effect of ignoring the source contribution.\color{black}

In Table \ref{tab:FRB_sample}, we report the sample of \color{black}65 \color{black} FRBs used in this work along with their redshifts and the observed DM. 
In Table \ref{tab:FRB_sample}, we give the ISM contribution for the FRBs which depends on their sky 
location and computed primarily using NE2001 model \citep{CL2002} (although see Sec. \ref{subsec:ISM} for results using the YMW2017 ISM model). These values can be found in the literature such as \cite{Zhang2023,WM2023,Law2024}. We also use a value of $\rm {DM_{halo}}=50$ pc cm$^{-3}$ which is typically assumed in the literature \citep{Macquart2020}
(see Sec. \ref{app:halo}) for results using ${\rm DM_{halo}}=100$ pc cm$^{-3}$ ). We note that this contribution is uncertain and can have consequences for inferred quantities such as $H_0$ (Hubble parameter today) as we discuss in Sec \ref{sec:halo_deg}. 
\begin{table}[H]
  \begin{center}
   \begin{tabular}{l|c|c|c|c|r} 
   FRB & z & $\rm {DM_{obs}}$ & $\rm {DM_{ISM}}$ (NE2001) &  $\rm {DM_{ISM}}$ (YMW2017) & References \\
    \hline
   FRB20220319D & 0.0112 & 110.98 & 133.3 & 210.96 & \cite{Law2024}  \\
   FRB20180916B & 0.0337 & 348.76 & 200 & 324.95 & \cite{Marcote2020}  \\
   FRB20220207C & 0.0430 & 262.38 & 79.3 & 83.27 & 
   \cite{Law2024}  \\
   FRB20211127I & 0.0469 & 234.83 & 42.5 & 31.46 &
   \cite{James2022}  \\
   FRB20201123A & 0.0507 & 433.55 & 251.93 & 162.4 & \cite{Rajwade2022} \\
   FRB20190303A & 0.064 & 222.4 & 26 & 21.79 & \cite{Michilli2023} \\
   FRB20210405I & 0.066 & 565.17 & 516.1 & 348.7 & \cite{Driessen2024} \\
   FRB20180814 & 0.068 & 189.4 & 87 & 108 & \cite{Michilli2023}
   \\
   FRB20231120A & 0.07 & 438.9 & 43.8 & 36.22 & \cite{Connor2024} \\
   FRB20220912A & 0.0771 & 219.46 & 115 & 122.24 & \cite{Ravi2023} \\
   FRB20220509G & 0.0894 & 269.53 & 55.2 & 52.06 &
   \cite{Law2024} \\
   FRB20230124 & 0.0940 & 590.6 & 38.5 & 31.77 & \cite{Sharma2024},\cite{Connor2024} \\
   FRB20201124A & 0.098 & 413 & 123 & 196.67 & \cite{Ravi2022}  \\
   FRB20220914A& 0.1139 & 631.28 & 55.2 & 51.11 & \cite{Law2024} \\
   FRB20190608B & 0.1178 & 339 & 37 & 26.62 & \cite{Macquart2020} \\
   FRB20240213A & 0.1185 & 357.4 & 40.1 & 32.10 & \cite{Connor2024} \\
   FRB20230628A & 0.1265 & 345.15 & 39.1 & 33.36 & \cite{Sharma2024},\cite{Connor2024} \\
   FRB20210410D & 0.1415 & 578.78 & 56.2 & 42.2 & \cite{Caleb2023} \\
   FRB20220920A & 0.158 & 314.99 & 40.3 & 30.83 & \cite{Law2024} \\
   FRB20200430A & 0.1608 & 380.25 & 27 & 26.08 & \cite{Heintz2020} \\
   FRB20210603A & 0.177 & 500.15 & 40 & 30.79 & \cite{Cassanelli2024} \\
   FRB20240215A & 0.21 & 549.5 & 48.0 & 42.79 & \cite{Connor2024} \\
   FRB20210117A & 0.214 & 729.0 & 34.0 & 23.0 & \cite{Bhandari2023} \\
   FRB20221027A & 0.229 & 452.5 & 47.2 & 40.59 & \cite{Connor2024} \\
   FRB20191001A & 0.234 & 506.92 & 44.7 & 31.08 & \cite{Heintz2020}\\
   FRB20190714A & 0.2365 & 504.13 & 38 & 31.16 & \cite{Heintz2020} \\
   FRB20221101B & 0.2395 & 490.7 & 131.2 & 193.36 & \cite{Sharma2024},\cite{Connor2024} \\
   FRB20220825A & 0.2414 & 651.24 & 79.7 & 86.98 & \cite{Law2024} \\
   FRB20191228A & 0.2432 & 297.5 & 33 & 19.93 & \cite{Bhandari2022}\\
   FRB20221113A & 0.2505 & 411.4 & 91.7 & 115.40 & \cite{Sharma2024},\cite{Connor2024} \\
   FRB20220307B & 0.2507 & 499.15 & 128.2 & 186.98 & \cite{Sharma2024},\cite{Connor2024} \\
   FRB20220831A & 0.262 & 1146.25 & 126.7 & 188 & \cite{Connor2024} \\
   FRB20231123B & 0.2625 & 396.7 & 40.2 & 33.81 & \cite{Sharma2024},\cite{Connor2024} \\
   FRB20230307A & 0.2710 & 608.9 & 37.6 & 29.47 & \cite{Sharma2024},\cite{Connor2024} \\
   FRB20221116A & 0.2764 & 640.6 & 132.3 & 196.18 & \cite{Connor2024} \\
   FRB20221012A & 0.2846 & 441.08 & 54.4 & 50.55 & \cite{Law2024} \\
   FRB20240229A & 0.287 & 491.15 & 37.9 & 29.52 & \cite{Connor2024} \\
   FRB20190102C & 0.2913 & 363.6 & 57.3 & 43.38 & \cite{Macquart2020} \\
   FRB20220506D & 0.3004 & 396.97 & 89.1 & 72.83 & \cite{Law2024} \\
   FRB20230501A & 0.3010 & 532.5 & 125.6 & 180.17 & \cite{Sharma2024},\cite{Connor2024} \\
   FRB20180924C & 0.3214 & 361.42 & 40.5 & 27.66 & \cite{Bannister2019} \\
   FRB20230626A & 0.3270 & 451.2 & 39.2 & 32.51 & \cite{Sharma2024},\cite{Connor2024} \\
   FRB20180301A & 0.3304 & 536 & 152 & 253.96 & \cite{Bhandari2022}\\
   
   \end{tabular}
   \end{center}
   \end{table}
   \begin{table}[H]
  \begin{center}
   \begin{tabular}{l|c|c|c|c|r} 
   FRB & z & $\rm {DM_{obs}}$ & $\rm {DM_{ISM}}$ (NE2001) &  $\rm {DM_{ISM}}$ (YMW2017) & References \\
    \hline
     FRB20231220A & 0.3355 & 491.2 & 49.9 & 44.54 & \cite{Connor2024} \\
   FRB20211203C & 0.3439 & 635.0 & 63.4 & 48.37 & \cite{Gordon2023_1} \\
   FRB20220208A & 0.3510 & 437.0 & 101.6 & 128.80 & \cite{Sharma2024},\cite{Connor2024} \\
   FRB20220726A & 0.3610 & 686.55 & 89.5 & 111.40 & \cite{Sharma2024},\cite{Connor2024} \\
   FRB20200906A & 0.3688 & 577.8 & 36 & 37.87 & \cite{Bhandari2022}\\
   FRB20240119A & 0.37 & 483.1 & 37.9 & 30.98 & \cite{Connor2024} \\
   FRB20220330D & 0.3714 & 468.1 & 38.6 & 30.09 & \cite{Sharma2024},\cite{Connor2024} \\
   FRB20190611B & 0.3778 & 321.4 & 57.8 & 43.67  &\cite{Heintz2020} \\
   FRB20220204A & 0.4 & 612.2 & 50.7 & 46.03 & \cite{Connor2024} \\
   FRB20230712A & 0.4525 & 586.96 & 39.2 & 30.93 & \cite{Sharma2024}, \cite{Connor2024} \\
   FRB20181112A & 0.4755 & 589.27 & 42 & 29.03 & \cite{Prochaska2019} \\
   FRB20220310F & 0.4779 & 462.24 & 45.4 & 39.51 &\cite{Law2024} \\
   FRB20190711A & 0.5220 & 593.1 & 56.4 & 42.62 &\cite{Heintz2020} \\
   FRB20230216A & 0.5310 & 828.0 & 38.5 & 27.05 & \cite{Sharma2024}, \cite{Connor2024} \\
   FRB20230814A & 0.5535 & 696.4 & 104.9 & 134.83 & \cite{Connor2024} \\
   FRB20221219A & 0.5540 & 706.7 & 44.4 & 38.60 & \cite{Sharma2024}, \cite{Connor2024} \\
   FRB20190614D & 0.60 & 959.2 & 83.5 & 108.72 & \cite{Law2020} \\
   FRB20220418A & 0.6220 & 623.25 & 37.6 & 29.54 & \cite{Law2024} \\
   FRB20190523A &  0.6600 & 760.8 & 37 & 29.88 & \cite{Ravi2019}\\
   FRB20240123A & 0.968 & 1462.0 & 90.3 & 112.98 & \cite{Connor2024} \\
   FRB20221029A & 0.9750 & 1391.05 & 43.9 & 36.4 & \cite{Sharma2024,Connor2024} \\
   FRB20220610A & 1.01 & 1458.1 & 30.9 & 13.58 & \cite{Gordon2023} 
   \end{tabular}
   \caption{ List of FRBs used in this work along with their associated properties. The dispersion measures are in the units of pc cm$^{-3}$. We have one burst with ${\rm DM_{ISM}>DM_{\rm obs}}$ which is of course non-physical. We note that the estimate of ISM DM contribution in NE2001 \cite{CL2002} model uses a smooth distribution of Milky-way electrons as well as large scale fluctuations in their distribution. However, there is some discrepancy in the measurement of distances to high altitude pulsars using this model \citep{Chatterjee2009}. \color{black} We have also added a column with the ISM contribution computed from the modelling done by YMW2017 \cite{Yao2017}. \color{black}} 
   \label{tab:FRB_sample}
   \end{center}
   \end{table}

\subsection{IGM contribution to ${\rm DM_{obs}}$}
The IGM contribution can be expressed as \citep{Beniamini2021},
\begin{equation}
    <{\rm DM_{IGM}}>(z)=\frac{3cH_0\Omega_{\rm b0}}{8\pi Gm_{\rm p}}\int_0^z {\rm d}z' \frac{(1+z)\xi_e(z)}{\left[\Omega_{\rm m0}(1+z')^3+\Omega_{\Lambda 0}\right]^{1/2}}
    \label{eq:DM_IGM}
\end{equation}
where $\Omega_{\rm b0}, \Omega_{\rm m0}$ and $\Omega_{\Lambda0}$ are the fractional energy density of baryons, matter and dark energy compared to critical energy density today. We assume flat $\Lambda$CDM cosmology which ensures $\Omega_{\Lambda0}=1-\Omega_{\rm m0}$ at the redshifts of interest. We use the best fit value of these parameters as inferred from the CMB (Cosmic Microwave Background) experiments \citep{Planck2020}. The expression depends upon the electron fraction $\xi_e(z)$, which in turn depends upon the helium mass fraction $(Y)$. Using $Y\approx 0.25$, we have $\xi_e(z)\approx 0.87$ which holds at $z\lesssim 3$, when helium become fully ionized. We can also rewrite the Hubble parameter as $H_0=100h$ kms$^{-1}$Mpc$^{-1}$ and use the variable $h$ below. From Eq. \ref{eq:DM_IGM}, we find that ${\rm DM_{IGM}}\propto \Omega_{\rm b0}h$. Therefore, we can only constrain a degenerate combination of $\Omega_{\rm b0}h$ from FRB observations. However, CMB measurements can provide $\Omega_{\rm b0}h^2=0.022$ \citep{Planck2020} with a tight prior. Therefore, using this prior, we can break this degeneracy and, in that case, ${\rm DM_{IGM}}$ is inversely proportional to $h$ \citep{James2022}. 

While Eq. \ref{eq:DM_IGM} gives the expression for the mean IGM contribution, there can be significant scatter along different lines of sight as matter starts to clump in the late universe. This leads to a variance in the IGM contribution. This variance has been computed in the literature (see, e.g., \cite{Jaroszynski2019}) using cosmological simulations. It depends sensitively on baryonic physics and can potentially vary among different simulations. \color{black} In this work, we primarily assume the distribution to be lognormal and use the result of \citep{Ziegler2024} which were fit to the high resolution large volume CoDa (cosmic dawn) II cosmological simulation data \citep{CoDa2020}. The standard-deviation in $\log \mbox{DM}_{\rm IGM}$ is then
\begin{equation}
    \sigma_{\rm IGM}(z)={\rm sinh^{-1}}(0.316z^{-0.677}).
    \label{eq:sigma_IGM}
\end{equation}
We return to this choice of IGM DM modeling in Sec. \ref{sec:Illustris}.   \color{black}

\subsection{Contribution from host galaxy}
We assume the host galaxy contribution to be of the form,
\begin{equation}
    <{\rm DM_{host}}>={\rm DM_{ host,0}}(1+z)^{\alpha}
    \label{eq:DM_host}
\end{equation}
In most of earlier works, $\alpha$ is assumed to be zero. However, the authors of \cite{Zhang2020} reported an approximate value of $\alpha\approx 1$ from the IllustrisTNG simulation \citep{Nelson2019}. The number can depend sensitively on the type of galaxy. Recent zoom-in simulations such as  \cite{Beniamini2021,Orr2024} show that the host contribution can have a large scatter due to galaxy age, morphology and orientation around a mean value of $\sim 100$ pc cm$^{-3}$ (Fig. 6 of \cite{Beniamini2021} and Fig. 8 of \cite{Orr2024}). This scatter in DM$_{\rm host}$ may dominate over the IGM contribution especially at low redshifts. Similar to previous works, we have assumed the scatter of the host to be log-normal. Therefore, we rewrite Eq. \ref{eq:DM_host} as,
\begin{equation}
    \mu=\mu_0+\alpha {\rm log}(1+z)
    \label{eq:DM_host1}
\end{equation}
where $\mu={\rm log <{ DM_{host}}>}$ and $\mu_0={\rm log(DM_{ host,0})} $. We assume the scatter in host galaxy contribution to be, $\sigma_{\rm host}$=1, which was the best fit value obtained in \cite{Macquart2020}. \color{black} However, we will vary this parameter in Sec. \ref{app:sigma_host} and show that our results qualitatively do not change. The range of this parameter covers the suitable parameter space found in IllustrisTNG simulation \citep{Mo2023}, (Table 2 of the reference). In principle, $\sigma_{\rm host}$ may be $z$ dependent. However, since most of our FRBs in the sample are at $z\lesssim 0.4-0.5$, we do not expect our calculations to be sensitive to such a dependence. In the future, as we detect FRBs at higher redshifts, this aspect of the analysis may be refined. \color{black}

\begin{figure}
 \centering
 \includegraphics[width=\columnwidth]{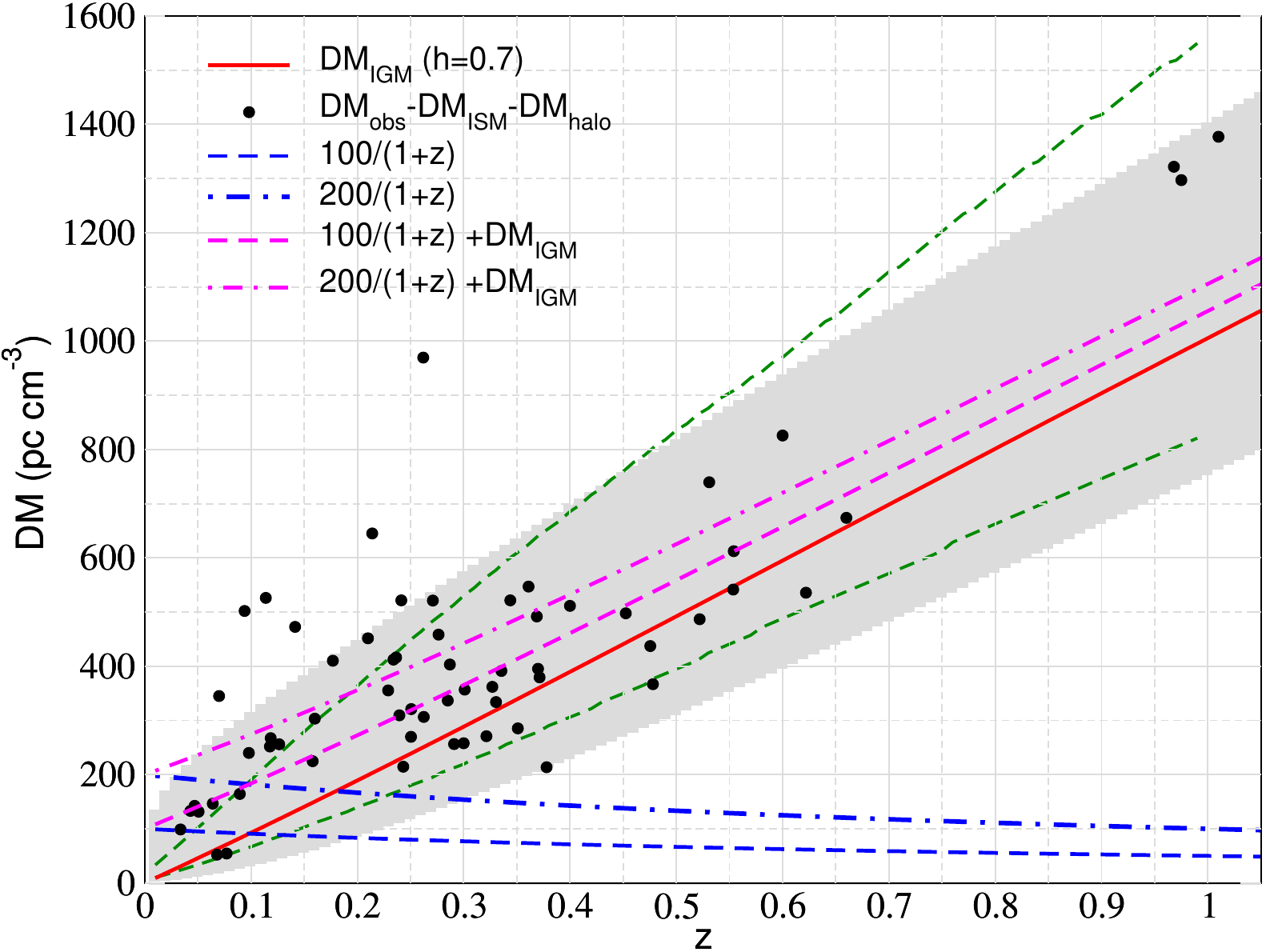}
 \caption{Dispersion measure of our sample of FRBs after subtracting ISM (NE2001 model) and halo contribution. We compare it to our fiducial IGM model with $h=0.7$ and host galaxy contribution with ${\rm DM_{host,0}}$=100 and 200 pc cm$^{-3}$, $\alpha=0$. The  shaded region around the red line shows the 1-$\sigma$ region due to scatter in IGM (Eq. \ref{eq:sigma_IGM}). At low redshifts, the host contribution dominates over IGM contribution which is shown visually by the blue double-sided arrow. \color{black}We also show an estimate of the 68 percent confidence interval of ${\rm DM_{IGM}}$ distribution obtained from IllustrisTNG simulation \citep{Zhang2021} in green dashed lines. \color{black}}
 \label{fig:sample1}
\end{figure}

In Fig. \ref{fig:sample1}, we plot the dispersion measure of our sample of FRBs after subtracting respective ISM and halo contribution. We compare this with a few cases assuming a fiducial cosmology and host galaxy contribution. We clearly see a preference for non-zero host contribution for sources at low redshifts where IGM contribution becomes sub-dominant. While the purpose of the figure is to visually motivate the case for non-zero host contribution, we will fit our model in Eq. \ref{eq:DM_host1}, directly to the data to extract the host galaxy contribution, below. 

\section{Likelihood analysis}
\label{sec:Likelihood}
\color{black}
In this work, we use Gaussian likelihood for analysis of data. For an individual FRB, the likelihood of the total observed ${\rm DM}$ is given by,
\begin{equation}
\mathcal{L}_i({\rm DM_i'}|z_i)=\int_0^{\rm DM_i'}\frac{{\rm  d}p_{\rm host}({\rm DM_{host}}|\mu,\sigma_{\rm host})}{{\rm d DM_{host}}}\frac{{\rm d}p_{\rm IGM}}{{\rm dlog} X}{\rm dDM_{host}},
\label{eq:likelihood}
\end{equation}
where, 
\begin{equation}
    \frac{{\rm d}p_{\rm host}}{{\rm dDM_{host}}}=\frac{1}{\sqrt{2\pi \sigma^2_{\rm host}}{\rm DM_{host}}}{\rm exp}\left(\rm -\frac{(log(DM_{host})-\mu)^2}{2\sigma^2_{\rm host}}\right),
\end{equation}
and,
\begin{equation}
    \frac{{\rm d}p_{\rm IGM}}{{\rm dlog} X}=\frac{1}{\sqrt{2\pi \sigma^2_{\rm IGM}}}{\rm exp}\left(\rm -\frac{[{\rm logX}-log(<DM_{IGM}>)]^2}{2\sigma^2_{\rm IGM}}\right),
    \label{eq:IGM_prob}
\end{equation}
with ${\rm DM_i'=DM_{i,obs}-DM_{i,ISM}-DM_{halo}}$ and $X={\rm DM_i'}-\frac{{\rm DM_{host}}}{1+z_i} $. Since all FRB sources are independent, the joint likelihood is given by the individual likelihood products,
\begin{equation}
    \mathcal{L}_{\rm tot}=\prod_i^N \mathcal{L}_i,
\end{equation}
where $N$ is the number of FRBs in the sample which is \color{black}65 \color{black} in our case. We use MCMC (Markov Chain Monte Carlo) sampling to scan over the parameter space. Up to this point the free parameters were $h,\mu_0$ and $\alpha$. However, the halo contribution is also uncertain and this can be degenerate with other terms such as the IGM contribution. Therefore, we study the degeneracy between $\rm DM_{halo}$ and $h$ first. \color{black}

\subsection{Degeneracy between halo contribution and $h$}
\label{sec:halo_deg}
\begin{figure}
 \centering
 \includegraphics[width=\columnwidth]{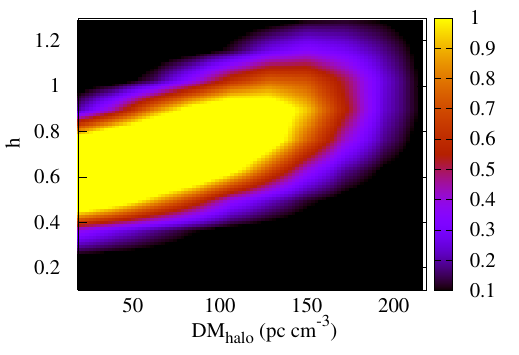}
 \caption{Joint probability $\frac{{\rm d}P}{\rm d(DM_{halo}h)}\times 100$ where the factor 100 is chosen for better normalization and plotting efficiency. }
 \label{fig:halo_h}
\end{figure}

We scan over $\rm DM_{halo}$ and $h$ in our likelihood analysis keeping $\alpha=0$ and $\mu_0$ to its reference value which is 100 pc cm$^{-3}$. We show the 2D posterior or correlation between the two parameters in Fig. \ref{fig:halo_h}. As we increase $\rm DM_{halo}$, the contribution from the IGM has to decrease in order to compensate. As we have shown previously, ${\rm DM_{IGM}}\propto h^{-1}$, the value of $h$ has to increase.   \color{black} This explains the behaviour seen in Fig. \ref{fig:halo_h}. Therefore, one needs to vary both ${\rm DM_{halo}}$ and $h$ simultaneously for consistency. This increases the dimensionality of the problem. In the literature, various authors have fixed ${\rm DM_{halo}}$ and scanned over $h$ in order to infer its value from FRB data \citep{Macquart2020,HR2022}. This potentially leads to biased results as Fig. \ref{fig:halo_h} shows. This can be a problem once we have a large number of FRBs such that the error bar on the inferred value of $h$ shrinks appreciably. For simplicity, we have fixed both ${\rm DM_{halo}}$ and $h$ to their assumed fiducial values of 50 pc cm$^{-3}$ and 0.7 respectively. Therefore, in \S \ref{Sec:results}, we have two remaining free parameters $\mu_0$ and $\alpha$ which we vary to fit the data.

\section{Results}
\label{Sec:results}

\begin{figure}[!htp]
\begin{subfigure}[b]{0.4\textwidth}
\includegraphics[scale=0.3]{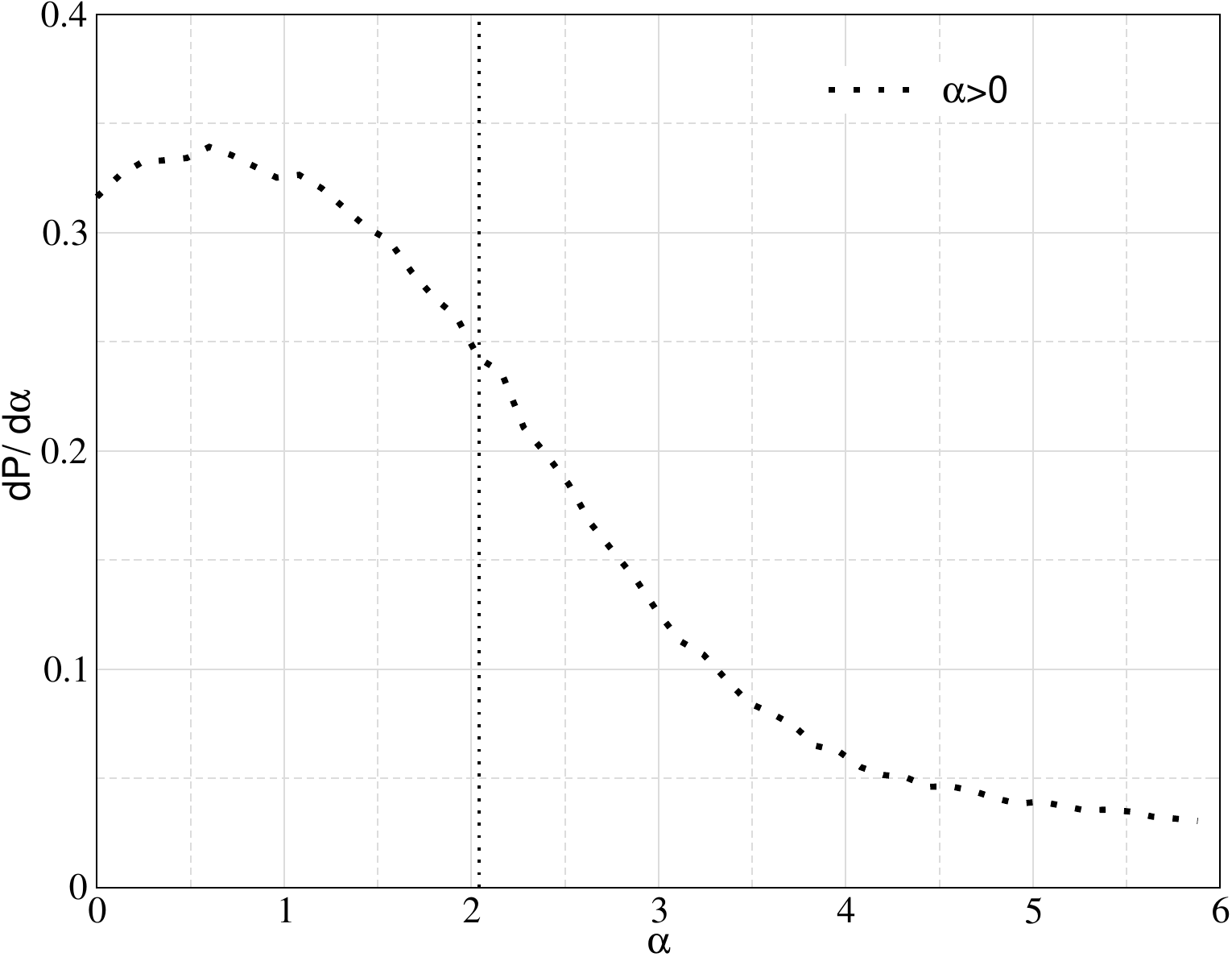}
\end{subfigure}\hspace{50 pt}
\begin{subfigure}[b]{0.4\textwidth}
\includegraphics[scale=0.3]{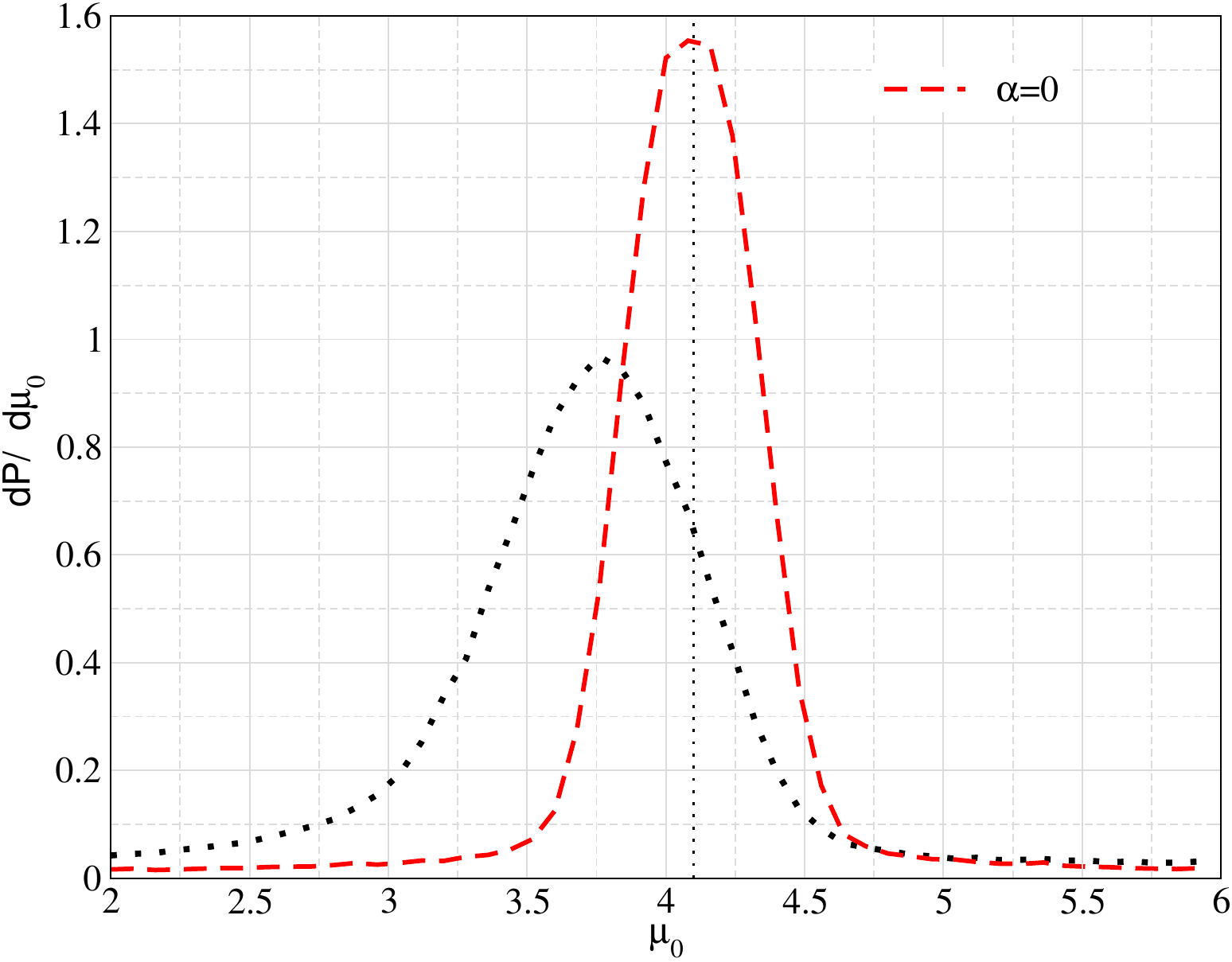}
\end{subfigure}
\caption{Probability distribution function of $\alpha$ and $\mu_0$ obtained by putting a prior on $\alpha>0$. For our fiducial case with $\alpha=0$, the best fit value of $\mu_0\approx 4$ which amounts to ${\rm DM_{host,0}}\approx 50$ pc cm$^{-3}$. \color{black}We show 68 percent confidence interval for the case when either $\alpha$ or $\mu_0$ are varied in vertical lines. In the black curve, we fit both $\alpha$ and $\mu_0$ to the data. }
 \label{fig:constraints}
\end{figure}

\begin{figure}[!htp]
\begin{subfigure}[b]{0.4\textwidth}
\includegraphics[scale=0.3]{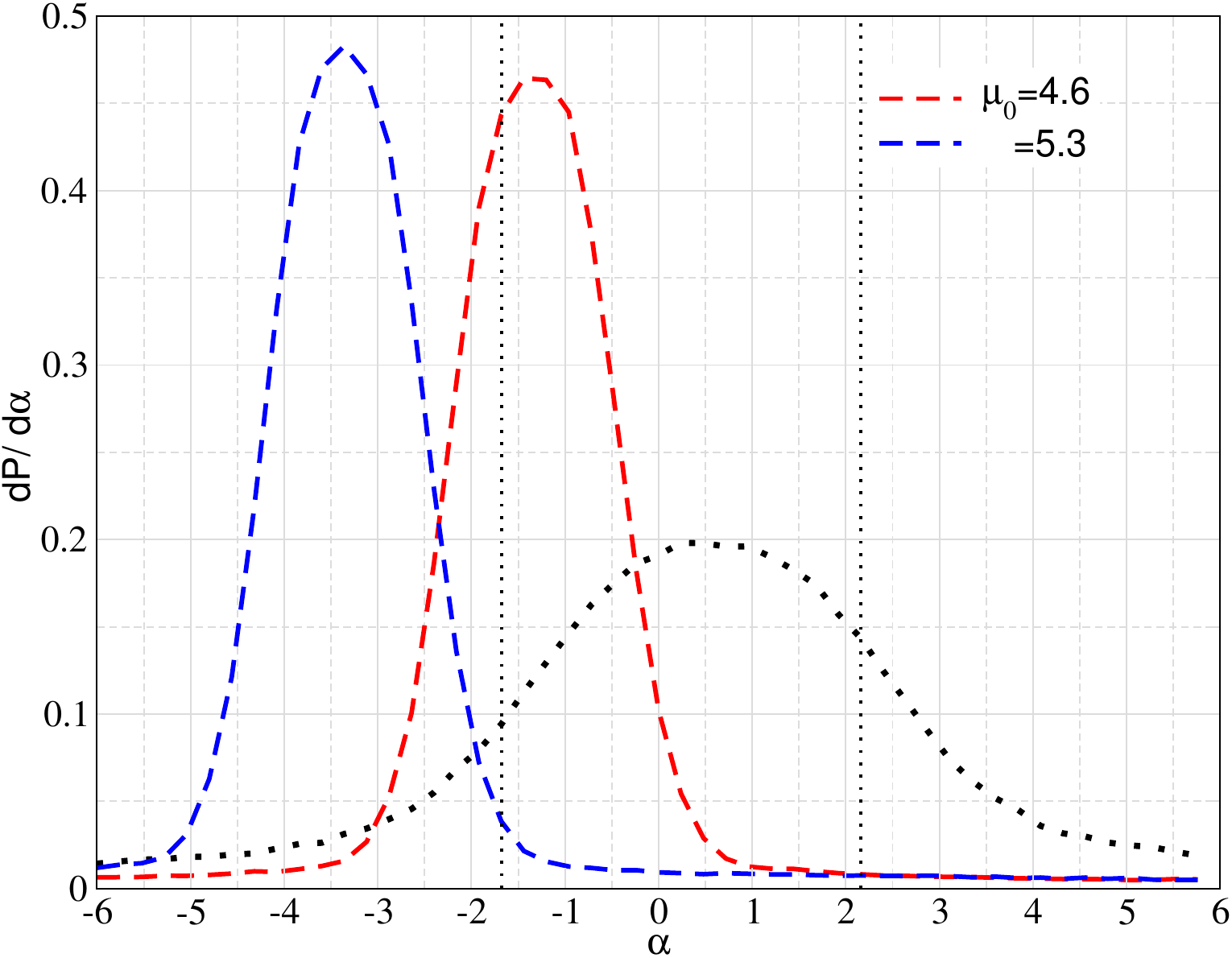}
\end{subfigure}\hspace{50 pt}
\begin{subfigure}[b]{0.4\textwidth}
\includegraphics[scale=0.3]{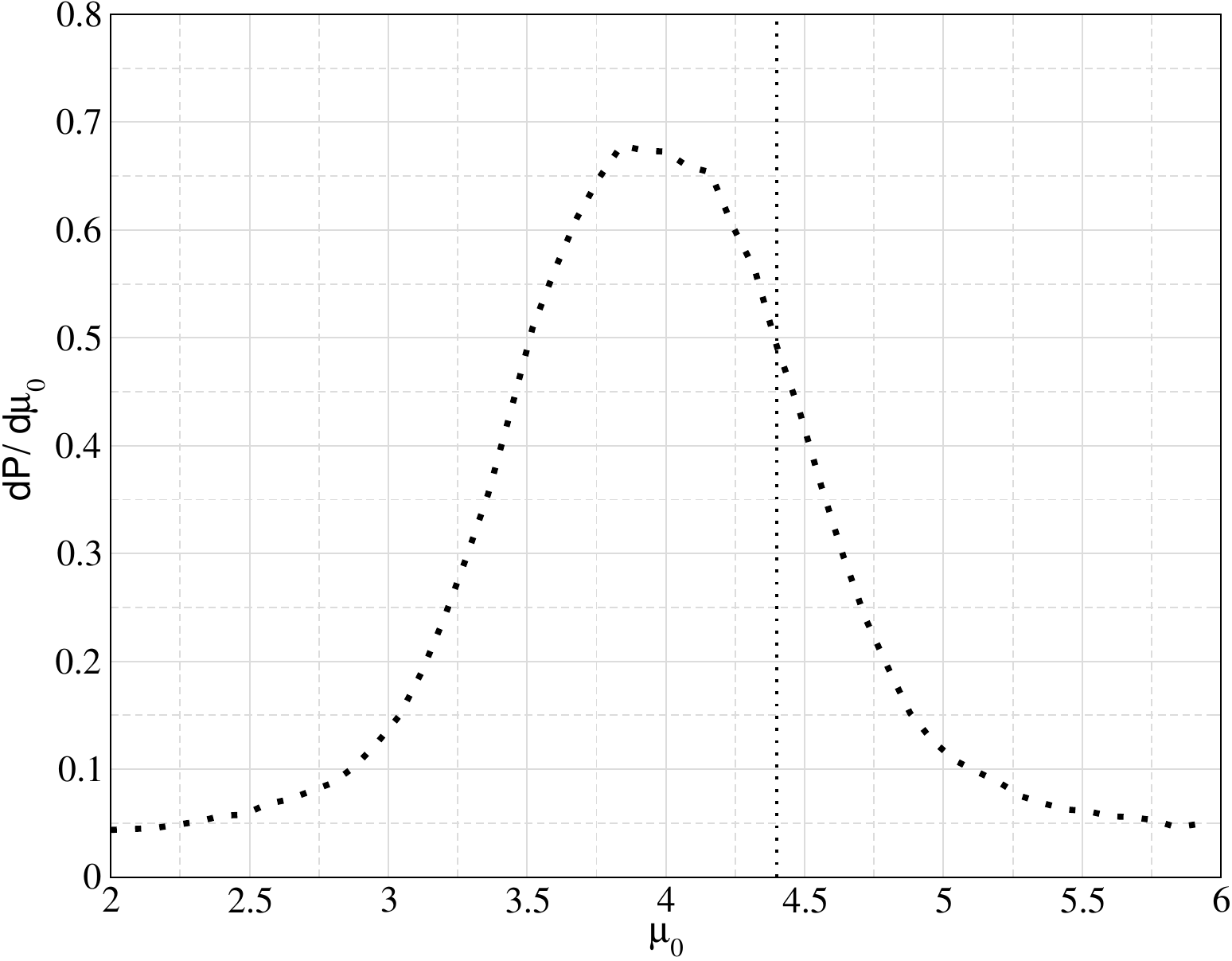}
\end{subfigure}
\caption{Constraints on $\alpha$ and $\mu_0$ by allowing $\alpha$ to have either sign. We consider two cases with fixed $\mu_0$ with best fit value such that ${\rm e}^{\mu_0}\approx 100$ pc cm$^{-3}$ and 200 pc cm$^{-3}$ respectively. In the black curve, we fit both $\alpha$ and $\mu_0$ to the data.}
 \label{fig:constraints1}
\end{figure}

In Fig. \ref{fig:constraints} and \ref{fig:constraints1}, we plot the main results of this work. We show the 1D posterior on each parameter after marginalizing over the other parameters. In Fig. \ref{fig:constraints}, we impose the criteria that $\alpha>0$. This is motivated from the fact that at higher redshifts, the universe is denser. Therefore, due to higher density, the host contribution should increase with redshift (see e.g. estimates by \cite{Beniamini2021} of DM$_{\rm host}$ at high $z$ based on zoom-in FIRE simulations) . We find that the current data can already put interesting constraints on $\alpha$. We find that $\alpha=0$ is preferred over higher values and that $\alpha\gtrsim 2$ can be ruled out a 68\% confidence. \color{black}For our fiducial case with $\alpha=0$, the best fit value of $\mu_0\approx 4$ or ${\rm DM_{host,0}}\approx 50$ pc cm$^{-3}$. For $\alpha>0$ case, the best fit value is even smaller but still comparable to the results obtained in \cite{Macquart2020}. These numbers are sensitive to the choice of $\sigma_{\rm IGM}$. This is expected since for a wider ${\rm DM_{IGM}}$ distribution the observed DM can be entirely explained by the IGM contribution.  \color{black} 

As a next step, we do not impose the positivity constraint on $\alpha$ and allow it to have both signs. The results are plotted in Fig. \ref{fig:constraints1}. \color{black}In such a case, $\alpha\approx 0.24$ has the highest likelihood \color{black}. The best fit value of $\mu_0$ remains about $\approx 4$. We find that a high value of $\alpha$ (with either sign) is strongly disfavoured. In the future, as our sample size increases, the upper limit will also reduce approximately as $\sqrt{N}$ where $N$ is the number of FRBs in the sample. Thereby, by increasing our sample size, we may rule out $\alpha\gtrsim 1$ at high statistical significance and, therefore, we should be in a position to test the results of \cite{Zhang2020,Kovacs2024}. Even without additional data, if we impose a strong prior on $\mu_0$, we can already obtain very strong constraints. As shown in Fig. \ref{fig:constraints1}, by fixing $\mu_0$ to a few representative values, there is a strong preference for negative $\alpha$ which is expected because of the degeneracy (Eq. \ref{eq:DM_host}). As FRBs are detected at higher and higher redshifts, the constraints should tighten as the degeneracy between $\alpha$ and $\mu_0$ will be broken increasingly more effectively.

\subsection{Comparison with a different ISM model}
\label{subsec:ISM}
\begin{figure}[!htp]
\begin{subfigure}[b]{0.4\textwidth}
\includegraphics[scale=0.3]{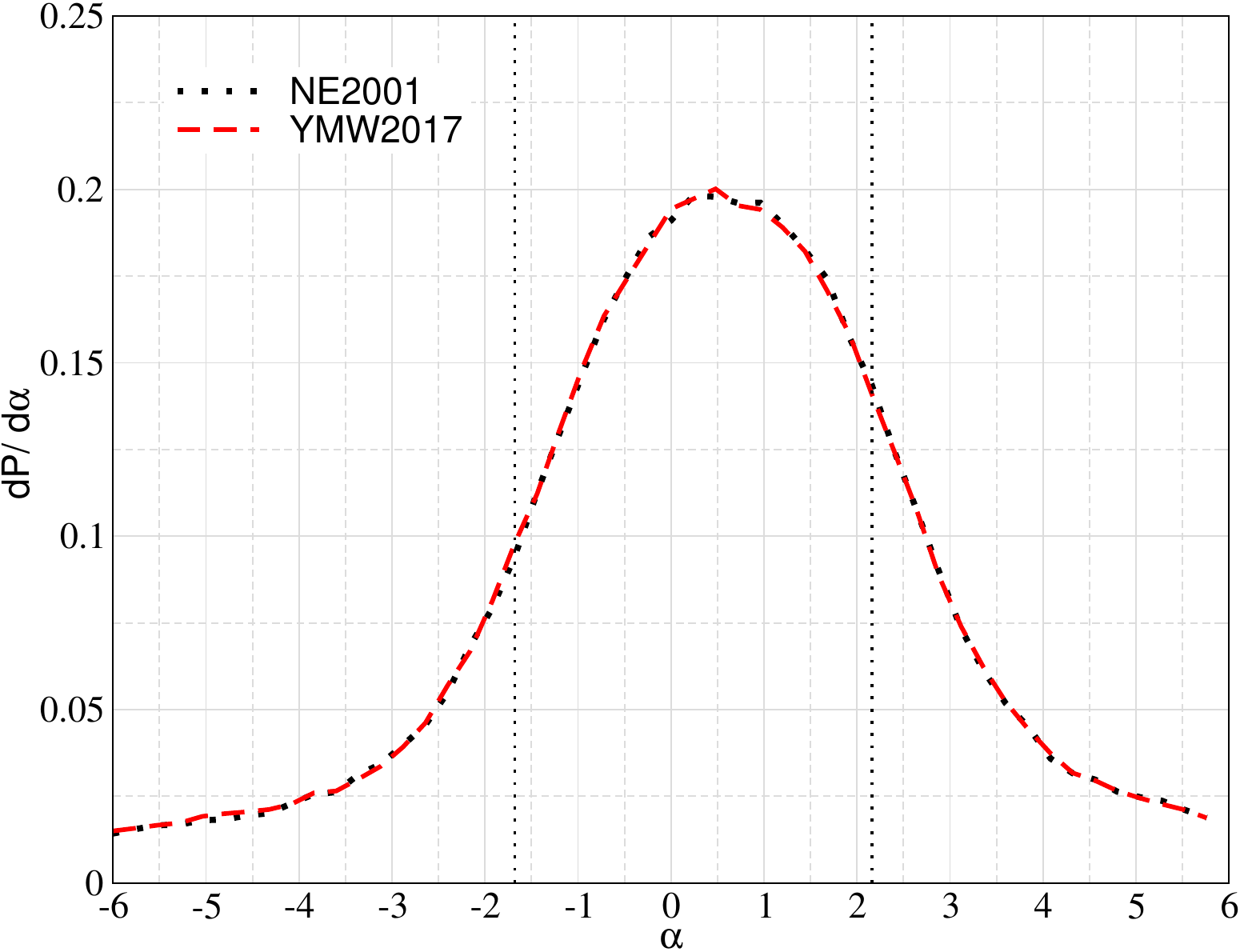}
\end{subfigure}\hspace{50 pt}
\begin{subfigure}[b]{0.4\textwidth}
\includegraphics[scale=0.3]{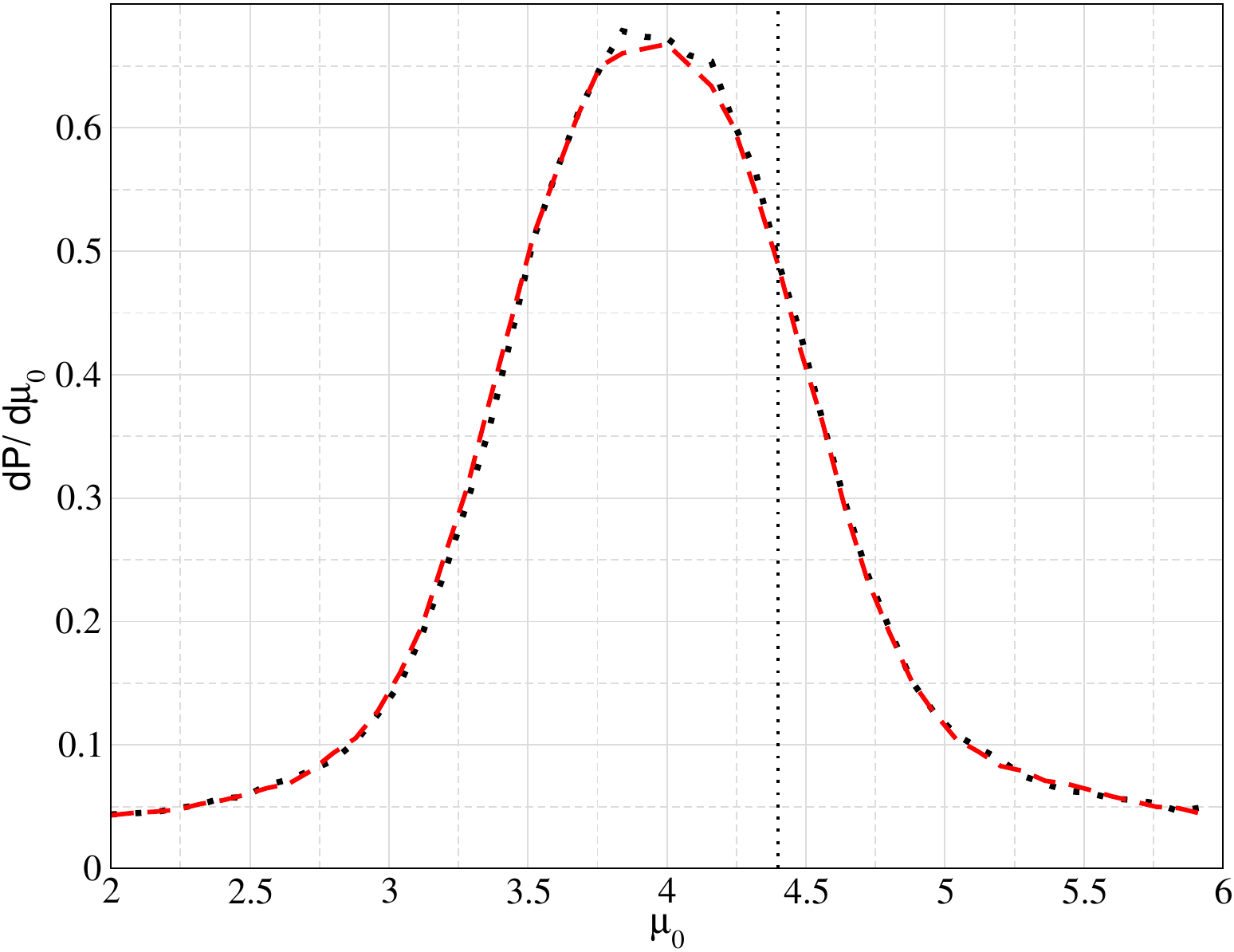}
\end{subfigure}
\caption{Constraints on $\alpha$ and $\mu_0$ comparing the two different ISM models. \color{black} We show the confidence interval for only one case since the the two cases produce roughly identical results. \color{black}}
 \label{fig:constraint_ISM}
\end{figure}

\color{black} In order to evaluate the uncertainty introduced by the model chosen for computing the ISM contribution, we compare our results with both the NE2001 \citep{CL2002} and YMW2017 \citep{Yao2017} models (shown in Table \ref{tab:FRB_sample}) \footnote{https://www.atnf.csiro.au/research/pulsar/ymw16/index.php?}. We find that the uncertainty in modelling does not introduce significant changes to our constraint as seen in Fig. \ref{fig:constraint_ISM}. In Sec. \ref{app:halo}, we check for uncertainty introduced by a different halo contribution.\color{black}

\subsection{Comparison with $\sigma_{\rm IGM}$ obtained from Illustris simulation}
\label{sec:Illustris}
\begin{figure}[!htp]
\begin{subfigure}[b]{0.4\textwidth}
\includegraphics[scale=0.3]{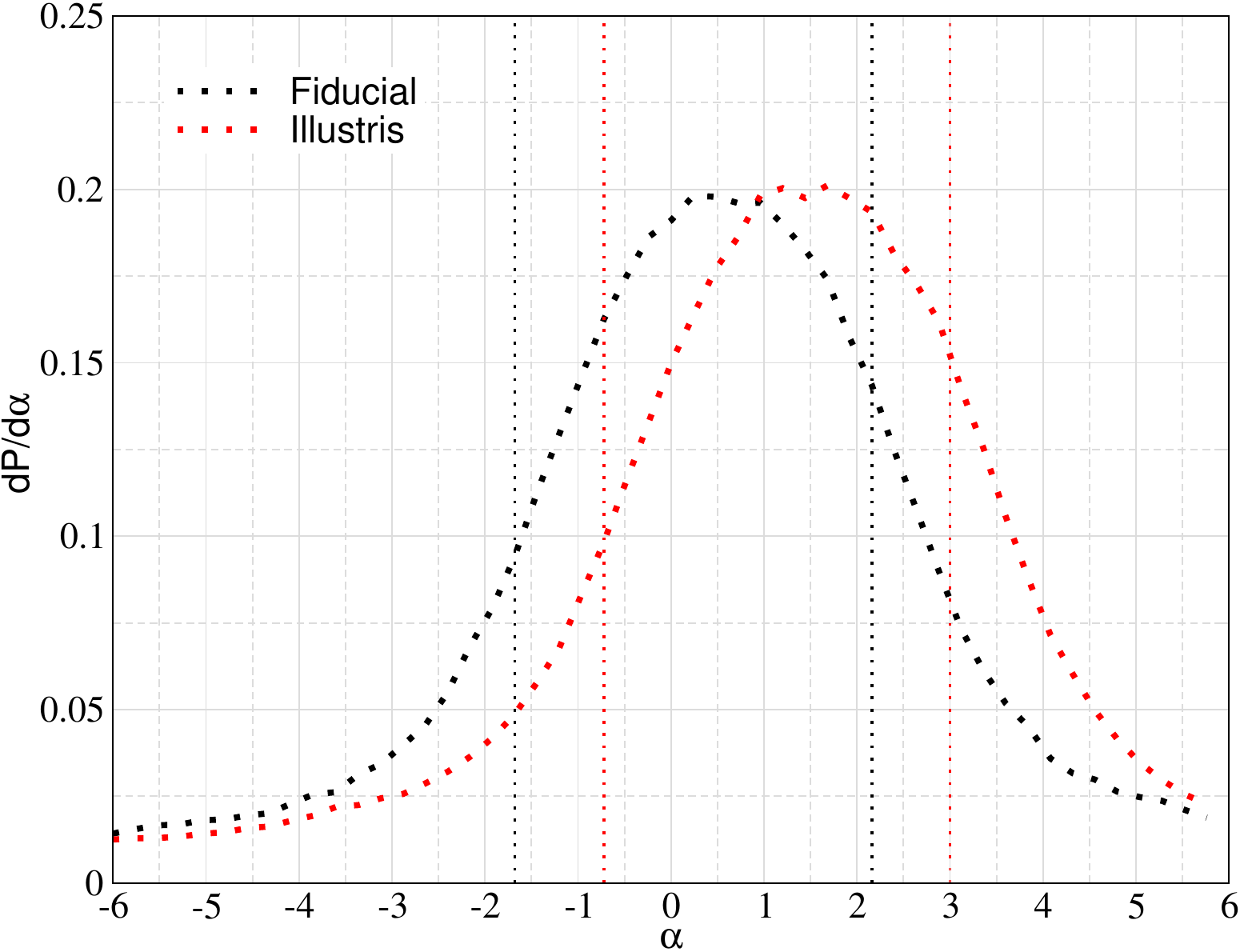}
\end{subfigure}\hspace{50 pt}
\begin{subfigure}[b]{0.4\textwidth}
\includegraphics[scale=0.3]{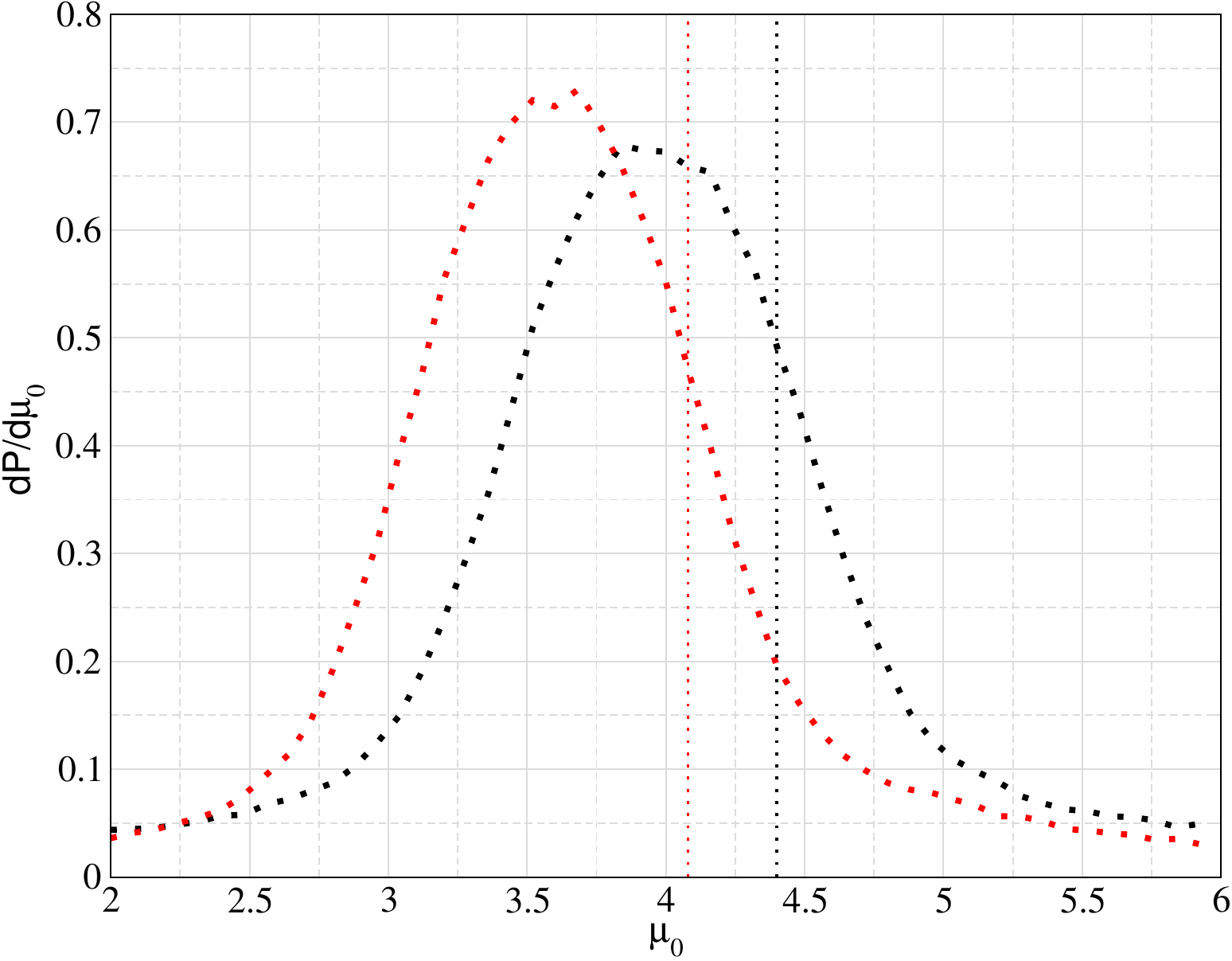}
\end{subfigure}
\caption{Constraints on $\alpha$ and $\mu_0$ using the results from IllustrisTNG simulation \citep{Zhang2021}. }
 \label{fig:constraints_illustris}
\end{figure}

\color{black}
In this work, we have assumed $\sigma_{\rm IGM}$ to be of the form in Eq. \ref{eq:sigma_IGM}. From Fig. \ref{fig:sample1}, we expect our constraints on $\alpha$ to be sensitive to the choice of $\sigma_{\rm IGM}$ since the observed DM has to be attributed to either IGM or the host contribution once the halo and ISM contribution have been accounted for. To showcase this sensitivity, we redo one of our computations using the result of the IllustrisTNG simulation. The DM distribution is fitted with the expression \citep{Zhang2021},
\begin{equation}
    \frac{{\rm d}p_{\rm IGM}}{{\rm d DM_{IGM}}}=A\Delta^{-\Tilde{\beta}}{\rm exp}\left[-\frac{(\Delta^{-\Tilde{\alpha}}-C_0)^2}{2\Tilde{\alpha}^2\sigma^2}\right]
\end{equation}
where $\Delta=\frac{{\rm DM_{IGM}}}{{\rm<DM_{IGM}>}}$, $\Tilde{\alpha}=\Tilde{\beta}=3$ and the fitting values of $A, C_0$ and $\sigma$ are provided in \citep{Zhang2021}. The equivalent expression for Eq. \ref{eq:IGM_prob} is given by,
\begin{equation}
   \frac{{\rm d}p_{\rm IGM}}{{\rm dlog}X}=AX\Delta^{-\Tilde{\beta}}{\rm exp}\left[-\frac{(\Delta^{-\Tilde{\alpha}}-C_0)^2}{2\Tilde{\alpha}^2\sigma^2}\right],
\end{equation}
where $X={\rm DM'}-\frac{{\rm DM_{host}}}{1+z}$. Then, we can use this expression in Eq. \ref{eq:likelihood} to compute the likelihood. We show our results in Fig. \ref{fig:constraints_illustris}. We find constraints which prefer slightly positive value of $\alpha\approx 1$ but otherwise are consistent with our fiducial result. 
The best fit value of $\mu_0$ also shifts to lower values which shows the degeneracy between $\alpha$ and $\mu_0$.  \color{black}

\subsection{Degeneracy between inferred $\Omega_{\rm b0}h$ and $\alpha$}
\begin{figure}[!htp]
\begin{subfigure}[b]{0.4\textwidth}
\includegraphics[scale=0.3]{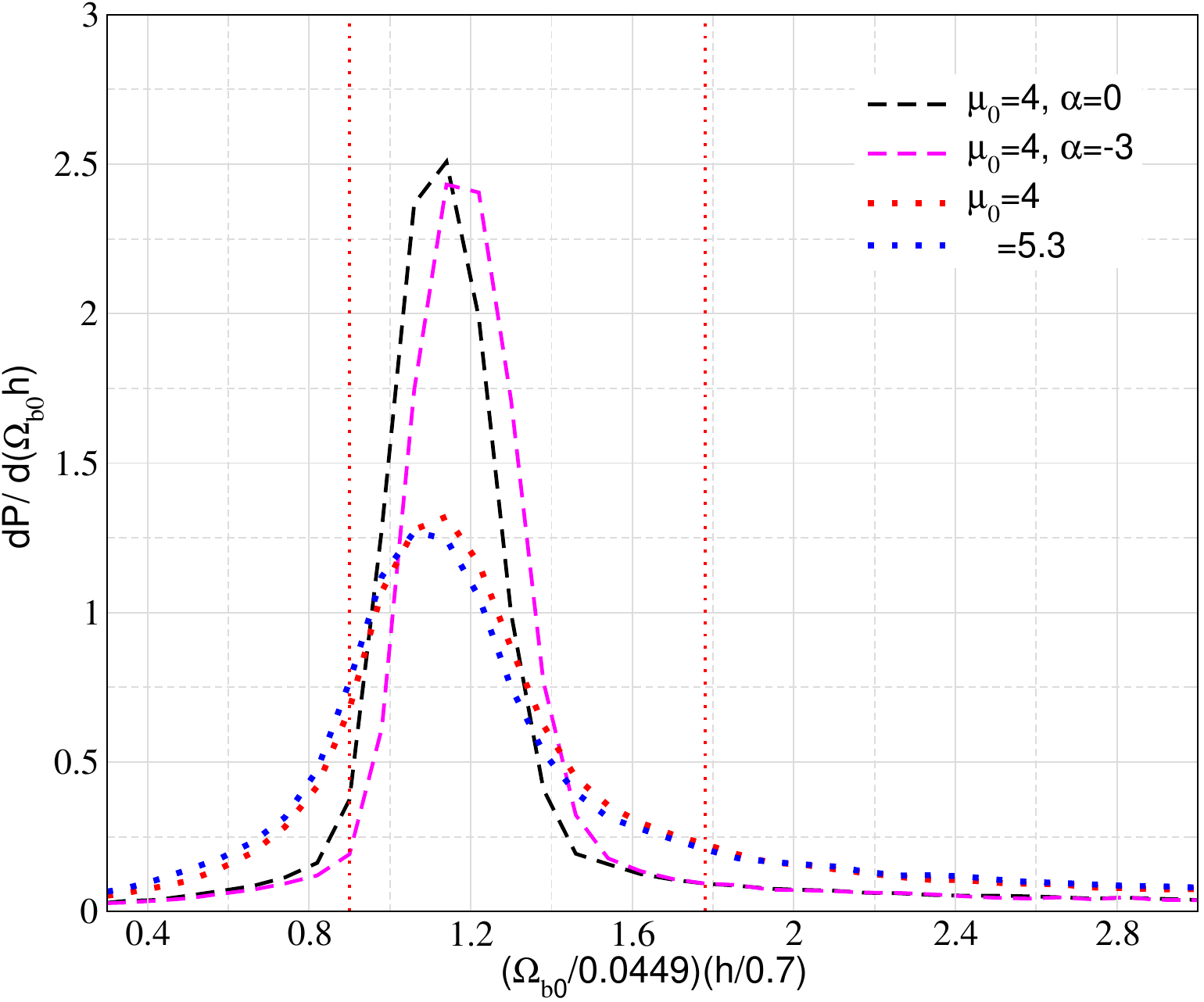}
\end{subfigure}\hspace{50 pt}
\begin{subfigure}[b]{0.4\textwidth}
\includegraphics[scale=0.3]{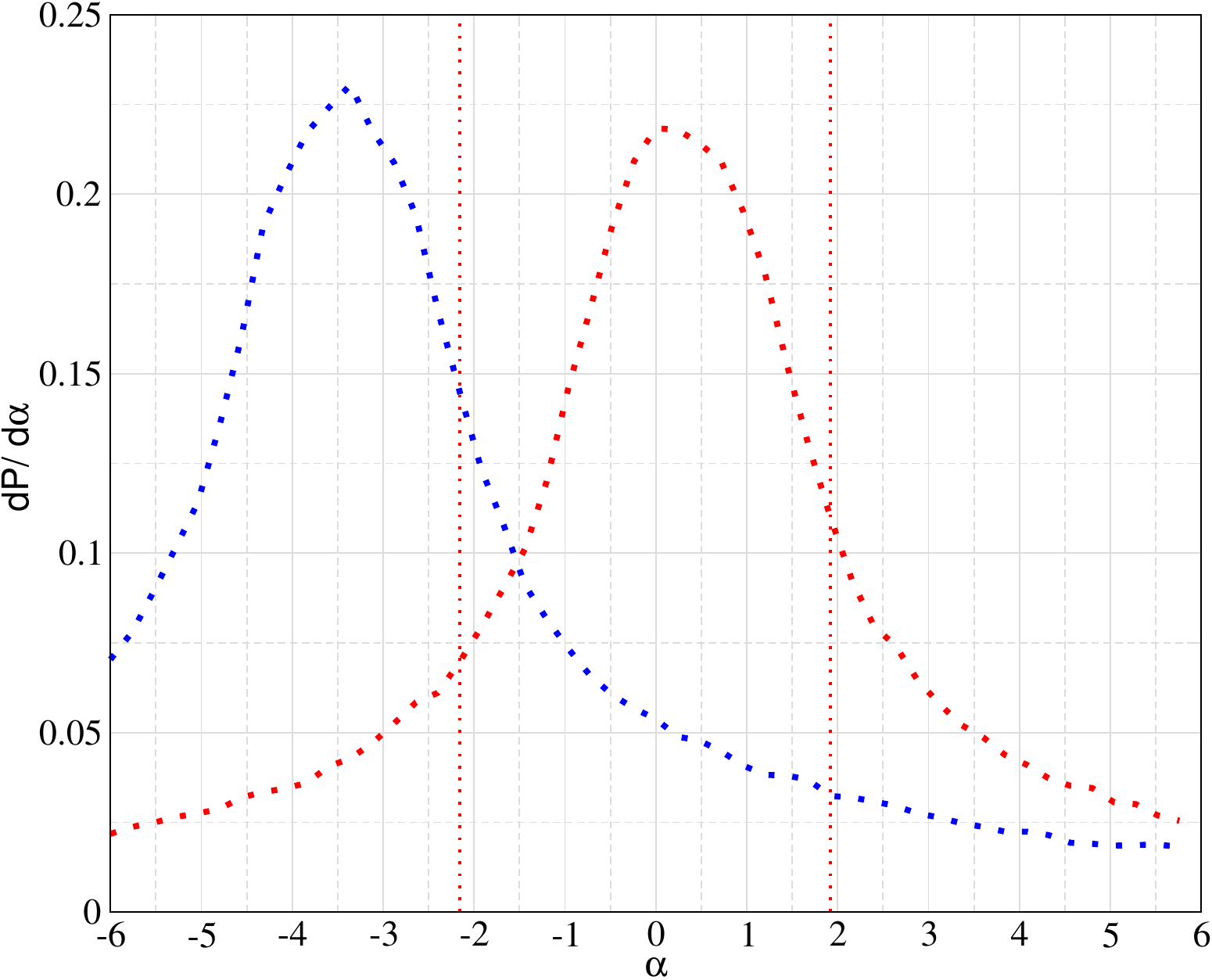}
\end{subfigure}
\caption{Constraints on $\Omega_{\rm b0}h$ and $\alpha$ by fitting these two parameters to data while keeping $\mu_0$ fixed to its prescribed value. We derive constraints on $\Omega_{\rm b0}h$ for two cases, where we fix $\alpha$ (dashed lines) as well as $\mu_0$ and two cases (dotted lines) where we fix $\mu_0$ while marginalizing over $\alpha$. }
 \label{fig:constraints2}
\end{figure}

Next, we consider the degeneracy between the inferred cosmological parameters $\Omega_{\rm b0}h$ and $\alpha$. We use Eq. \ref{eq:DM_IGM} to compute the ${\rm DM_{IGM}}$ contribution and do not use CMB prior to eliminate $\Omega_{\rm b0}$. We also fix ${\rm DM_{halo}}$ to its assumed value and $\mu_0$ as denoted in Fig. \ref{fig:constraints2}. The best fit value of $\Omega_{\rm b0}h$ does not shift much and $\alpha$ roughly has a similar distribution to that shown in Fig. \ref{fig:constraints1}. This is expected since $\Omega_{\rm b0}h$ is just a constant while $\alpha$ controls the redshift dependence of the host galaxy DM contribution. Therefore, there isn't a significant correlation between the two. However, there is a non-negligible $\Omega_{\rm b0}h$ distribution tail at both ends. This small degeneracy can be broken as more FRBs are detected and particularly, ones at higher and higher redshifts. 

\section{Discussion and conclusions}
\label{Sec:conclusions}
We have constrained the redshift dependence of host galaxy dispersion measures assuming that the dependence can be parameterized by a power law in $(1+z)$. We considered a sample of \color{black}65 \color{black} FRBs with redshifts up to $\approx 1$ and fitted the observed DM with the combination of host galaxy, IGM, Milky way and its halo contribution. Using the current data, the parameter space with \color{black} $\alpha\in$ [0 -- 1] is preferred with best fit value of $\alpha=0.24^{+1.92}_{-1.92}$. Depending on our modelling choices, we obtain $\alpha\lesssim 2$ at 68 percent confidence. \color{black} These results can inform us about galaxy formation physics and how the host DM depends on it. There is a significant degeneracy between $\alpha$ and $\mu_0$ which compensate each other for the host galaxy contribution. \color{black}Since $\alpha$ captures the redshift dependence while $\mu_0$ is the host contribution at $z\approx 0$, we expect observed DM to be more sensitive to changes in $\alpha$ at higher redshifts. Therefore, we expect this degeneracy to be broken as we detect FRBs at higher redshifts. \color{black}

The host DM can also have significant contributions from the circumgalactic medium. To a first approximation, this can be absorbed into our effective definition of DM$_{\rm host}$ (although it is unclear if the two terms should evolve with redshift in a similar way). Additionally, a foreground galaxy can contribute significantly to the DM of a background galaxy \citep{CR2022}, however, the fraction of such foreground galaxies appears to be $\lesssim 5\%$ (Fig. 1 of \cite{CR2022}). Therefore, we do not expect significant systematics to be introduced by ignoring these foregrounds. In the future, we may isolate a statistically significant sample of bursts whose lines of sight go through such foreground galaxies and repeat our analysis. These may have implications for our understanding on the distribution of diffuse gas around galaxies. \color{black} Recently, the authors in \cite{Connor2024} obtained an estimate on contribution coming from circumgalactic medium of intervening halos. They parameterize the DM from IGM contribution to two components, $f_X$ (clumpy matter component) and $f_{\rm IGM}$ (diffuse matter component) which are normalized so that they add up to the total matter in the universe. The authors use results from Illustris simulation extensively and obtain constraint on $f_X\approx 0.1$ using a sample of FRBs. In our work, we use the results from \cite{Ziegler2024} which captures the clumpy as well as diffuse component in a single object $\sigma_{\rm IGM}$. Therefore, we expect our results obtained with \cite{Ziegler2024} prescription to be conservative. In addition, we do not expect the results to be sensitive to the intervening halo contribution since the inferred $f_X$ is low. \color{black}

As a second step, we fit the data with varying $\Omega_{\rm b0}h$ and $\alpha$ for a few representative values of $\mu_0$. We find that the best fit value of $\Omega_{\rm b0}$ is stable to changes in $\alpha$, though, it adds a non-negligible amount of scatter around the best-fit $\Omega_{\rm b0}h$ value. We expect these degeneracies to be broken significantly with a bigger sample and more detected FRBs from higher redshifts. This is because the power law contribution $(1+z)^{\alpha}$ is sensitive to the available values of $z$. For the sample considered in this work, most of the FRBs are at $z\lesssim 0.4$ which makes the $(1+z)^{\alpha}$ term less sensitive to changes in redshift within the range stated above.  

Our results rely on a few assumptions. We have assumed knowledge of the scatter  in the host galaxy DMs, $\sigma_{\rm host}$. Alternatively, we may try to infer this directly from the data. We also point out that our inference of cosmological parameters such as $\alpha$ and $\Omega_{\rm b0}h$ depends sensitively on how we partition the observed DM to its various constituents. As an example, we show how the inferred value of $h$ changes as we vary ${\rm DM_{halo}}$ (Fig. \ref{fig:halo_h}). Ideally, we would want to fit all the unknown parameters to the data and obtain their posterior distribution. In the future, a large sample with a significant number of FRBs at $z\gtrsim 1$ will propel us towards that goal. Most detected FRBs, to date, do not have an identified redshift. This larger sample of FRBs has not been used in the present analysis. However, correlating this larger FRB sample with galaxy sky maps, can provide useful statistical constraints on the typical redshift of FRBs \citep{CHIME2021} even when redshifts of individual bursts remains undetermined and this information, in turn, can be fed back into the analysis presented in this work, to improve its statistical power.

\section*{Acknowledgements}
SKA is supported by the ARCO fellowship. PB is supported by a grant (no. 2020747) from the United States-Israel Binational Science Foundation (BSF), Jerusalem, Israel, by a grant (no. 1649/23) from the Israel Science Foundation and by a grant (no. 80NSSC 24K0770) from the NASA astrophysics theory program.

{
\vspace{-3mm}
\bibliographystyle{unsrtads}
\bibliography{main}
}

\appendix

\section{Dependence of constraints on $\sigma_{\rm host}$}
\label{app:sigma_host}
\begin{figure}[!htp]
\begin{subfigure}[b]{0.4\textwidth}
\includegraphics[scale=0.3]{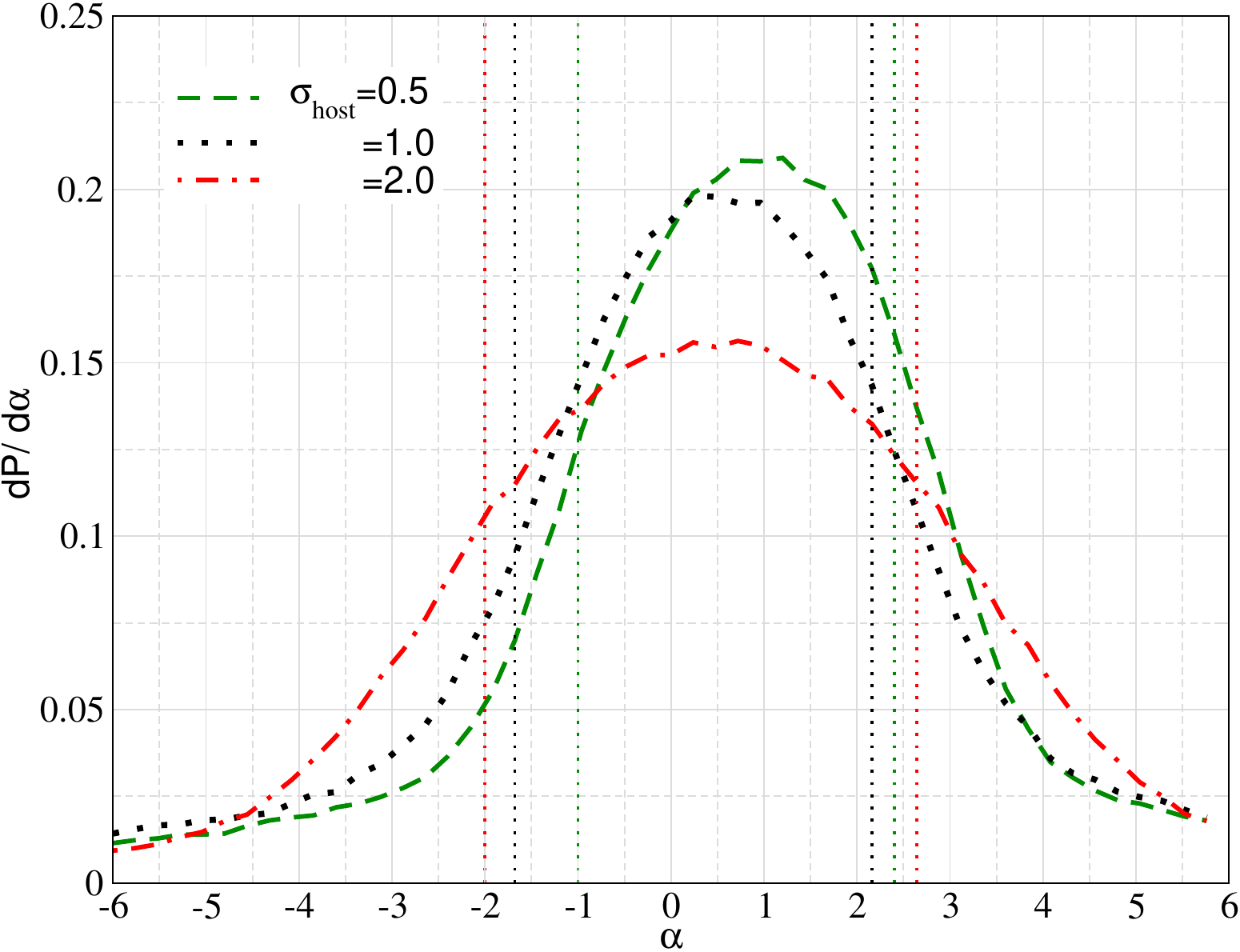}
\end{subfigure}\hspace{50 pt}
\begin{subfigure}[b]{0.4\textwidth}
\includegraphics[scale=0.3]{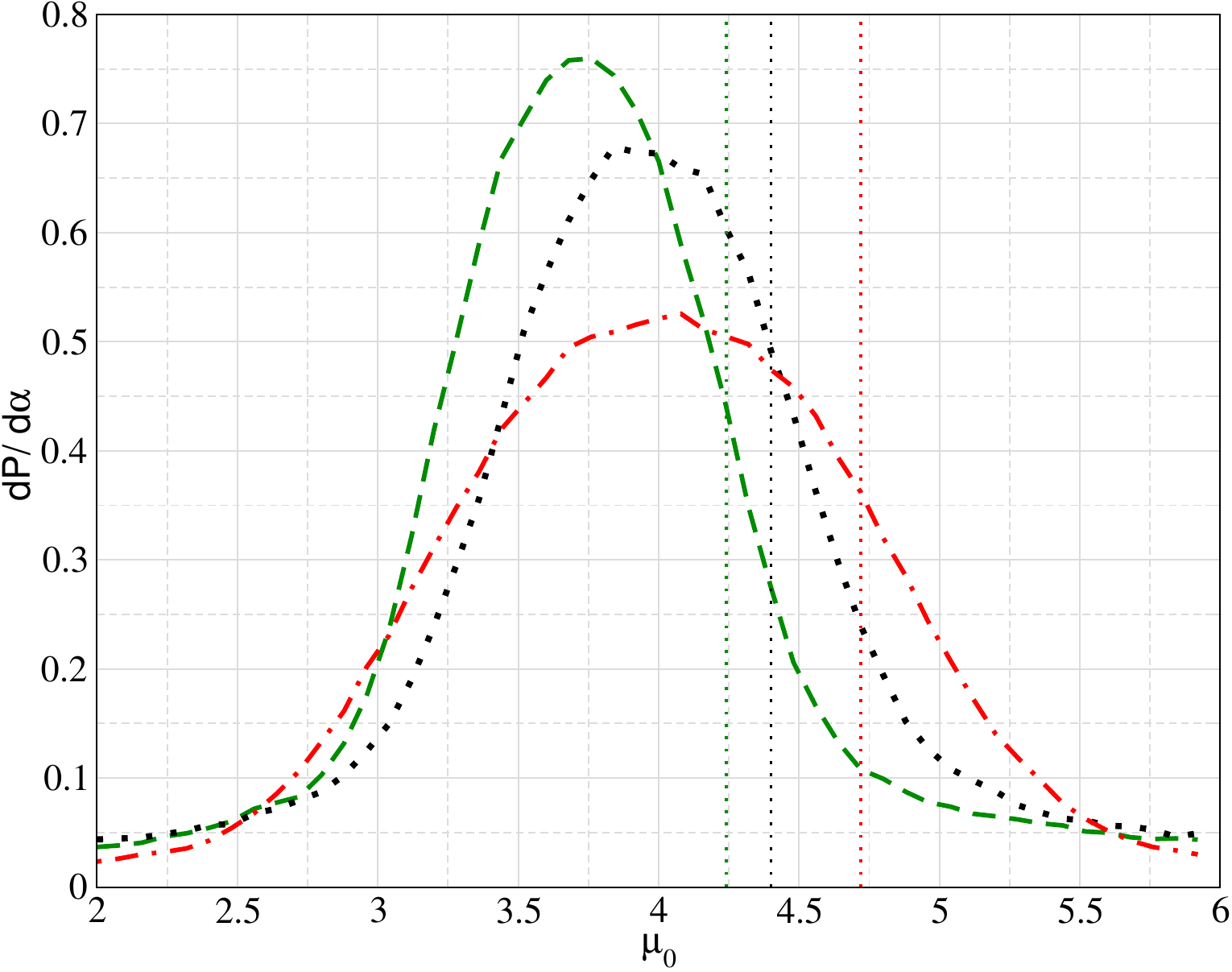}
\end{subfigure}
\caption{Constraints on $\alpha$ and $\mu_0$ by varying $\sigma_{\rm host}$.}
 \label{fig:constraints3}
\end{figure}

In Fig. \ref{fig:constraints3}, we vary $\sigma_{\rm host}$, the standard deviation of DM$_{\rm host}$.  Qualitatively, the constraints on $\alpha$ do not change drastically. However, the posterior distribution becomes flatter for higher $\sigma_{\rm host}$. In the future, one can try to infer $\sigma_{\rm host}$ from the data itself. For a small sample size as considered in this work, a larger number of parameters can lead to significant degeneracies. However, once we have a bigger sample, we may be able to break the degeneracies and infer these parameters consistently from the data.

\section{Degeneracy between $\alpha$ and ${\rm DM_{halo}}$}
\label{app:halo}
\begin{figure}[!htp]
\begin{subfigure}[b]{0.4\textwidth}
\includegraphics[scale=0.3]{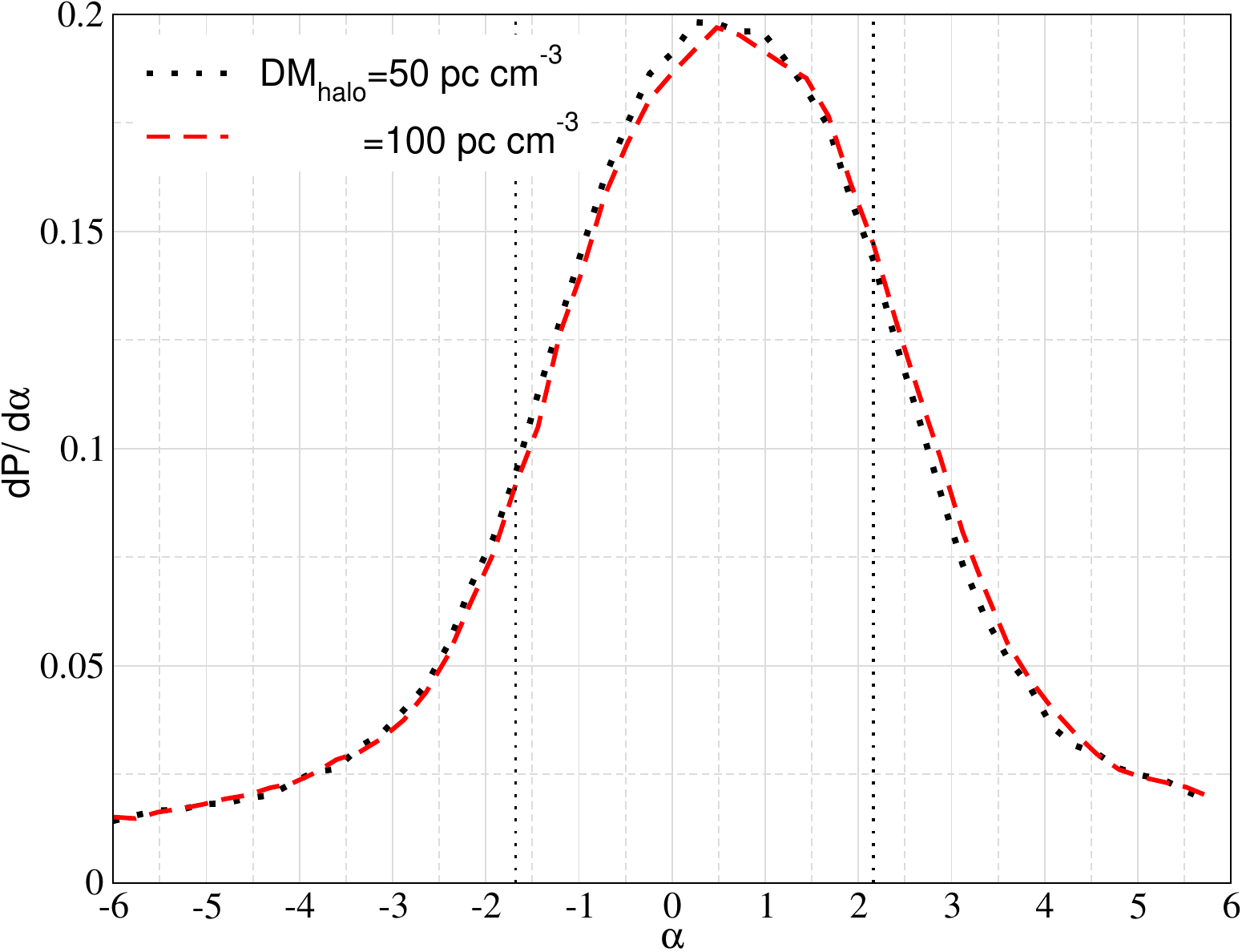}
\end{subfigure}\hspace{50 pt}
\begin{subfigure}[b]{0.4\textwidth}
\includegraphics[scale=0.3]{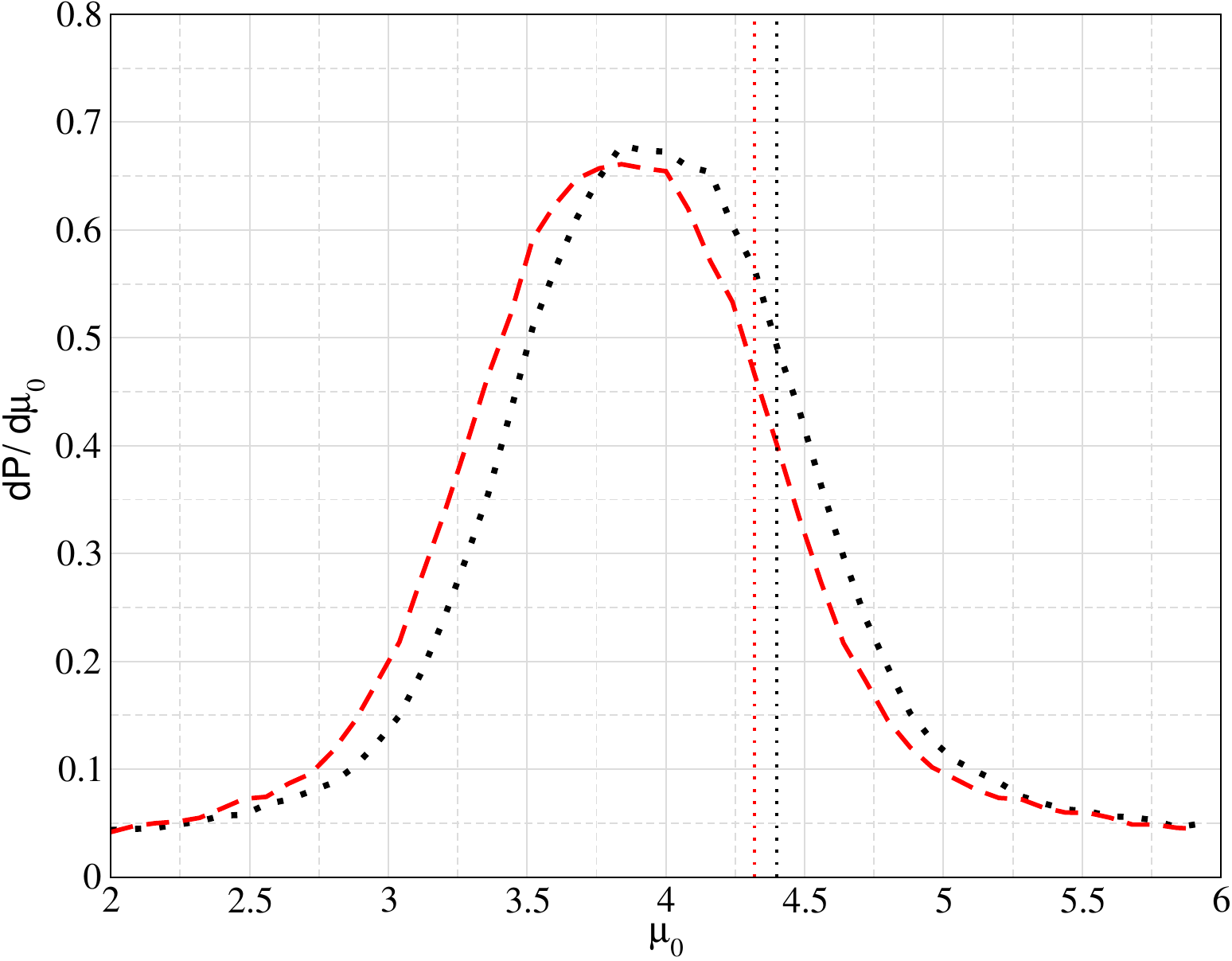}
\end{subfigure}
\caption{Constraints on $\alpha$ and $\mu_0$ for a couple of values of ${\rm DM_{halo}}$. }
 \label{fig:constraints_halo}
\end{figure}

In this work, we have fixed the value of ${\rm DM_{\rm halo}}$ to 50 pc cm$^{-3}$. However, there can be large uncertainty in the halo contribution \citep{PZ2019,YT2020}. In Fig. \ref{fig:constraints_halo}, we study the variation on constraints on $\alpha$ by changing ${\rm DM_{halo}}$ to 100 pc cm$^{-3}$. We do not find any significant degeneracy between $\alpha$ and ${\rm DM_{halo}}$. This is to be expected since ${\rm DM_{halo}}$ is a redshift-independent number while $\alpha$ captures the redshift dependence. Instead, the change in ${\rm DM_{halo}}$ is accommodated by a reduction in the value of $\mu_0$ which accounts for the host contribution at $z=0$.  \color{black}  

\end{document}